# On the Quantum Corrections to the Newtonian Potential

*H.W. Hamber and S. Liu*

Physics Department
University of California at Irvine
Irvine, Ca 92717

## ABSTRACT

The leading long-distance quantum correction to the Newtonian potential for heavy spinless particles is computed in quantum gravity. The potential is obtained directly from the sum of all graviton exchange diagrams contributing to lowest non-trivial order to the scattering amplitude. The calculation correctly reproduces the leading classical relativistic post-Newtonian correction. The sign of the perturbative quantum correction would indicate that, in the absence of a cosmological constant, quantum effects lead to a slow increase of the gravitational coupling with distance.

May 1995

# 1   Introduction

It is generally assumed that a quantum theory of gravity cannot lead to testable predictions, due to a lack of perturbative renormalizability of the Einstein-Hilbert action [1, 2, 3, 4, 5]. Recently the interesting possibility has been raised [6] that low energy predictions of quantum gravity are not necessarily affected by the short distance details of an ultraviolet regulated theory of gravity [7], and can in fact be finite and calculable. As an application, the leading long distance quantum correction to the static Newtonian potential was computed, resulting in a finite correction of order $O(G\hbar/c^3 r^3)$. When gravity is treated in this fashion as an effective low energy theory, the analogy with the use of effective field theories in treating the physics of soft pions comes to mind [8].

The existence of a universal long distance quantum correction to the Newtonian potential should be relevant for a wide class of gravity theories. It is known that the ultraviolet behavior of pure Einstein gravity can be improved by adding higher derivative contributions to the action [7]. In four dimensions these can be restricted to the form $\alpha R^{\mu\nu} R_{\mu\nu} + \beta R^2$, where $\alpha$ and $\beta$ are dimensionless coupling constants. The resulting classical and quantum corrections to gravity are expected to alter significantly the potential at short distances (comparable to the Planck length) but should not affect the behavior at large distances, which should largely be determined by the structure of the Einstein-Hilbert action. Only the latter action will be therefore the subject of our present investigation. We should add that simplicial lattice regularizations of gravity also include in general higher derivative terms, and the same considerations should apply in this case as well, as long as the correct continuous gauge invariance properties of the continuum action are incorporated [9].

In the following we will compute the leading classical and quantum corrections to the static potential, by evaluating the *complete* set of diagrams which contribute to the scattering amplitude for heavy spinless particles in the low momentum transfer limit. From the resulting expression the effective static potential at large distances can then be read off easily, and will contain, as explained further below, both classical relativistic and quantum corrections. An important omission in our calculation will be the absence of a bare cosmological constant, which would complicate the perturbative treatment significantly due to the need to expand about a non-flat background. Our results for the static potential are such that they suggest a slow increase of gravitational interactions with distance due to the quantum correction. The answer will then be compared with two recent calculations of the same quantity [6, 10, 11], which include only a subset of the diagrams considered here. We will find that our answer is qualitatively similar to the result of [6], but differs in sign from the result of [10], where a rather different method, based on world-line correlations, is used to estimate the potential.

## 2    One Loop Amplitudes

Before describing the calculation, it will be useful to first clarify our conventions and notation. We shall expand around the flat Minkowski space-time metric, with signature given by $\eta_{\mu\nu} = \text{diag}(1, -1, -1, -1)$. The Einstein-Hilbert action is then given by

$$S_{\text{E}} = +\frac{1}{16\pi G} \int dx \sqrt{-g(x)}\, R(x)\,, \tag{2.1}$$

with $g(x) = \det(g_{\mu\nu})$ and $R$ the scalar curvature. It is also assumed in the following that the bare cosmological constant is zero. The presence of a non-vanishing cosmological constant introduces additional momentum independent vertices, and would make the perturbative calculation described below considerably more difficult. In particular the expansion around flat space is no longer justified in this case, and it has to be performed around a solution of Einstein's equations with a non-zero cosmological constant.

The coupling of gravity to scalar particles of mass $m$ is described by the action

$$S_{\text{m}} = \frac{1}{2} \int dx \sqrt{-g(x)}\, \Big[\, g^{\mu\nu}(x)\partial_\mu\phi(x)\partial_\nu\phi(x)\, -\, m^2\phi^2(x)\,\Big] \tag{2.2}$$

In the following we shall consider the interaction induced by graviton exchange between two heavy scalar particles of distinct mass $m_1$ and $m_2$. The effective interaction in the static limit is then determined by evaluating the scattering amplitude between the two heavy particles, in the limit of small momentum transfer $\vec{q}^2 \to 0$.

Usually in perturbation theory the metric $g_{\mu\nu}(x)$ is expanded around the flat metric $\eta_{\mu\nu}$ [4], by writing

$$g_{\mu\nu}(x) = \eta_{\mu\nu} + \kappa\, \tilde{h}_{\mu\nu}(x) \tag{2.3}$$

with $\kappa^2 = 32\pi G$. Here we shall instead follow the method of reference [12], and define the small fluctuation graviton field $h_{\mu\nu}(x)$ via

$$g^{\mu\nu}(x)\sqrt{-g(x)} = \eta^{\mu\nu} + \kappa\, h^{\mu\nu}(x) \tag{2.4}$$

One advantage of this expansion over the previous one is that it leads to considerably simpler Feynman rules, both for the graviton vertices and for the scalar-graviton vertices (as can be seen from the fact that precisely the above expression appears in the kinetic term of the scalar field action). Once the action is expanded out in the graviton field $h^{\mu\nu}(x)$, the space-time indices are then raised and lowered using the flat metric, and there is therefore no longer a need to distinguish between upper and lower indices.

A gauge fixing term [13, 14] has to be introduced, and here it will be of the form

$$\frac{1}{\kappa^2}\left(\partial_\mu\sqrt{-g(x)}g^{\mu\nu}\right)^2, \tag{2.5}$$

as suggested in Ref. [12]. The bare graviton propagator is then given simply by

$$D_{\mu\nu\rho\sigma}(p) \;=\; \frac{i}{2}\,\frac{\eta_{\mu\rho}\eta_{\nu\sigma}+\eta_{\mu\sigma}\eta_{\nu\rho}-\eta_{\mu\nu}\eta_{\rho\sigma}}{p^2+i\epsilon}. \tag{2.6}$$

For the present calculation one also needs expressions for the three-graviton and two ghost-graviton vertex. The relevant expressions are quite complicated and have already been given in Ref. [12], so they will not be reproduced here. We have performed a number of checks of the results of Ref. [12], some of which will be discussed below. Let us point out here that with the present definition for the gravitational field, there are no factors of $1/(d-2)$ for the graviton propagator in $d$ dimensions; such factors appear instead in the expressions for the Feynman rules for the vertices. For the following calculations we shall also need the two scalar-one graviton vertex, which is given by

$$\frac{i\kappa}{2}\left(p_{1\mu}p_{2\nu}+p_{1\nu}p_{2\mu}-\frac{2}{d-2}\,m^2\,\eta_{\mu\nu}\right) \tag{2.7}$$

where the $p_1, p_2$ denote the four-momenta of the incoming and outgoing scalar field, respectively. In addition we need the two scalar-two graviton vertex, which is given by

$$\frac{i\kappa^2 m^2}{2(d-2)}\left(\eta_{\mu\lambda}\eta_{\nu\sigma}+\eta_{\mu\sigma}\eta_{\nu\lambda}-\frac{2}{d-2}\,\eta_{\mu\nu}\eta_{\lambda\sigma}\right) \tag{2.8}$$

where one pair of indices $(\mu,\nu)$ is associated with one graviton line, and the other pair $(\lambda,\sigma)$ is associated with the other graviton line. These rules follow readily from the expansion of the gravitational action to order $G^{3/2}$ $(\kappa^3)$, and of the scalar field action to order $G$ $(\kappa^2)$.

To lowest order in $G$, the contribution to the potential can be computed from the single graviton exchange diagram. In momentum space the static contribution is given, as expected, by

$$-\,G\,m_1\,m_2\,\frac{4\pi}{\vec{q}^2} \tag{2.9}$$

where $\vec{q}$ is the momentum transfer (see also [15]).

Higher order corrections in $G$ are computed by evaluating contributions to the interaction coming from the complete set of one-loop diagrams. One notices that the relevant length scale appearing with the Einstein-Hilbert action for pure gravity is the Planck length $l_p = (G\hbar/c^3)^{1/2}$. On the other hand the action for the scalar particle involves only the combination $mc/\hbar$, the inverse Compton wavelength associated with the heavy sources. This is also clearly seen from the path integral phase contribution for a single particle, which is given by

$$\frac{imc^2}{\hbar} \int_{\tau^{(a)}}^{\tau^{(b)}} d\tau \sqrt{g_{\mu\nu}(x(\tau))\frac{dx^\mu}{d\tau}\frac{dx^\nu}{d\tau}} \quad , \tag{2.10}$$

When one considers the lowest order contribution to the gravitational interaction due to single graviton exchange one obtains a contribution to the static gravitational potential proportional to $(\hbar/c)(mc^2/\hbar)^2(G\hbar/c^3) = m^2 G$. At order $G^2$ one finds contributions both of order $(\hbar/c)(mc^2/\hbar)^2(G\hbar/c^3)^2 = m^2\hbar G^2/c^3$ and of order $(\hbar/c)(mc^2/\hbar)^3(G\hbar/c^3)^2 = m^3 G^2/c$. The first one represents a genuine quantum correction proportional to $\hbar$, while in the second type of contribution the $\hbar$'s have canceled, and the resulting correction represents a classical relativistic correction. The latter involves the Schwarzschild radius of the massive particle, $2Gm/c^2$.

These considerations lead to the apparently paradoxical result that Feynman diagram perturbation theory is also expected to reproduce the classical relativistic corrections, which are independent of $\hbar$. Indeed it was shown by the authors of Ref. [16, 17] that classical relativistic corrections involve tree graphs connected to an arbitrarily high number of external classical sources. A detailed calculation of these classical relativistic corrections, using diagrammatic methods, was performed in Ref. [18]. There it was shown explicitly that the corrections of order $G^2$ correctly and completely reproduce the leading classical relativistic corrections appearing in the Einstein-Hoffmann-Infeld effective post-Newtonian Hamiltonian.

Let us now return to the computation of the one loop amplitude. One needs to calculate all first-order corrections in $G$, which will include both the classical-relativistic $O(G^2 m^2/c^2)$ and the quantum mechanical $O(\hbar G^2/c^3)$ corrections to the classical Newtonian potential energy discussed before. The relevant topologically distinct Feynman diagrams are shown in Figs. 1 and 2.

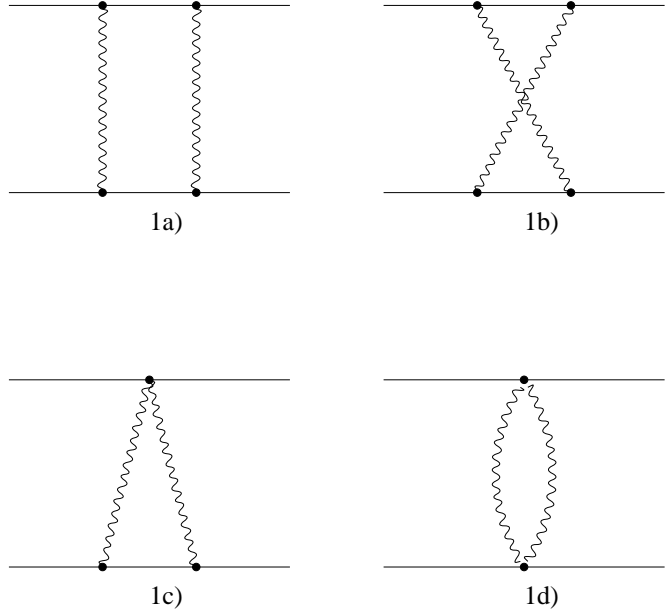

Fig 1. Some one loop graviton exchange diagrams.

The relevant amplitudes are computed in momentum space as a function of the total momentum transfer squared $\vec{q}^2$. They are then evaluated using dimensional regularization in $4 - \epsilon$ dimensions, using the Feynman parametric representation for combining propagator denominators. The final answer then follows after performing the necessary momentum and parametric integrations. Due to the vast amount of algebraic manipulations involved in doing the index contractions, computer algebra was employed throughout the calculation in order to ensure the correctness of the results. For small $\vec{q}^2$ the contributions arising from each diagram can then be separated into two types of terms, one describing the classical relativistic correction proportional to $1/\sqrt{\vec{q}^2}$, and the other describing the leading quantum correction proportional to $\log \vec{q}^2$.

These in turn can then be expressed as corrections in coordinate space by using

$$\int \frac{d^3\vec{q}}{(2\pi)^3} e^{-i\vec{q}\cdot\vec{x}} \frac{1}{\vec{q}^2} \to \frac{1}{4\pi r} \, . \tag{2.11}$$

$$\int \frac{d^3\vec{q}}{(2\pi)^3} e^{-i\vec{q}\cdot\vec{x}} \frac{1}{\sqrt{\vec{q}^2}} \to \frac{1}{2\pi r^2} \, . \tag{2.12}$$

$$\int \frac{d^3\vec{q}}{(2\pi)^3} e^{-i\vec{q}\cdot\vec{x}} \log \vec{q}^2 \to -\frac{1}{2\pi r^3} \, . \tag{2.13}$$

A nontrivial check of the calculation is then provided by the expected equality, for each diagram involving massless particles only, of the coefficient of the $2/\epsilon$ ultraviolet divergence and of the coefficient of the $-\log \vec{q}^2$ contribution, which would appear as one single logarithmic term $\log(\Lambda^2/\vec{q}^2)$ in the presence of an explicit ultraviolet cutoff $\Lambda$.

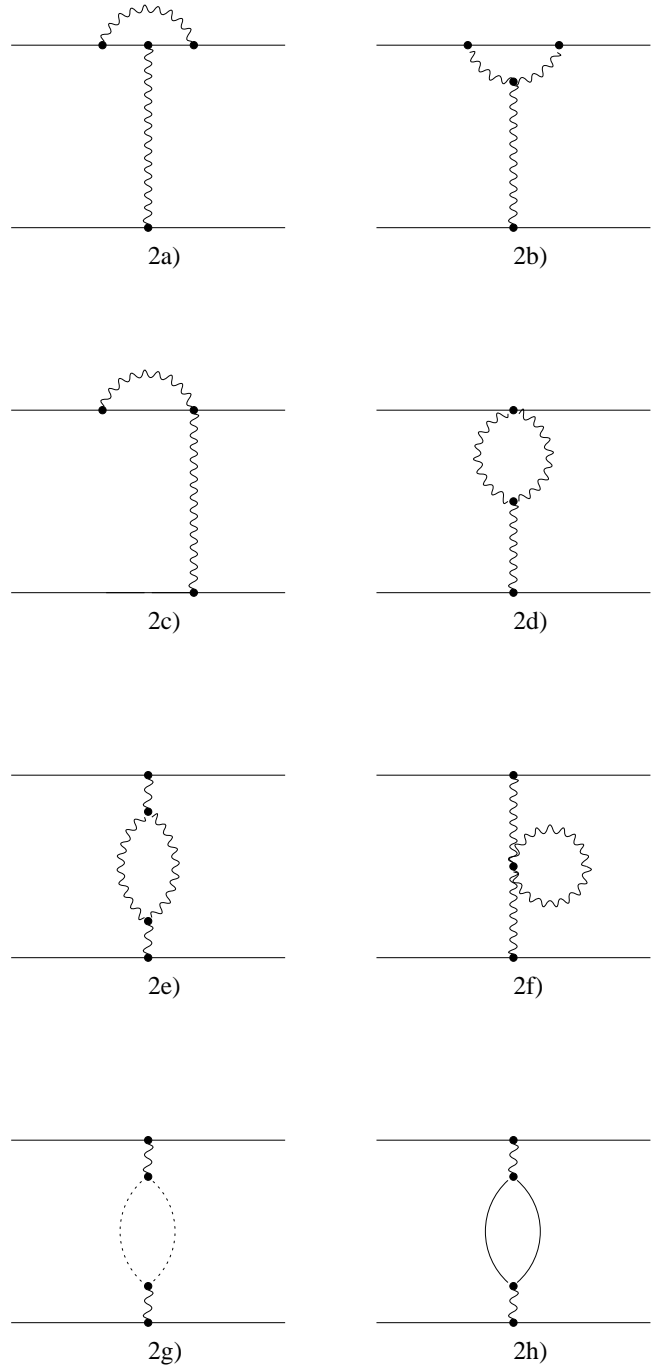

Fig 2. Additional one loop graviton exchange diagrams.

# 3  Results and Discussion

By converting the expressions for the individual diagrams to coordinate space, one obtains the following results. One has from diagram 1a

$$+\frac{3}{4}\,G^2\,\frac{m_1 m_2(m_1+m_2)}{r^2}+2\,G^2\,\frac{m_1 m_2}{\pi r^3},\tag{3.1}$$

from diagram 1b

$$+\frac{3}{4}\,G^2\,\frac{m_1 m_2(m_1+m_2)}{r^2}+2\,G^2\,\frac{m_1 m_2}{\pi r^3},\tag{3.2}$$

from diagram 1c

$$-\,G^2\,\frac{m_1 m_2(m_1+m_2)}{r^2}+8\,G^2\,\frac{m_1 m_2}{\pi r^3},\tag{3.3}$$

from diagram 1d

$$-\,10\,G^2\,\frac{m_1 m_2}{\pi r^3},\tag{3.4}$$

from diagram 2b

$$+\frac{16}{3}\,G^2\,\frac{m_1 m_2}{\pi r^3},\tag{3.5}$$

and from diagram 2d

$$+\frac{23}{3}\,G^2\,\frac{m_1 m_2}{\pi r^3}.\tag{3.6}$$

From diagrams 2e and 2g one obtains the graviton and ghost vacuum polarization contribution

$$-\frac{206}{30}\,G^2\,\frac{m_1 m_2}{\pi r^3}.\tag{3.7}$$

This last contribution was also computed in Ref. [12]. We have verified that the Slavnov-Taylor identity for the vacuum polarization $\Pi_{\alpha\beta\gamma\delta}(q)$,

$$q_\mu q_\nu\,D_{\mu\lambda\alpha\beta}(q)\,\Pi_{\alpha\beta\gamma\delta}(q)\,D_{\gamma\delta\nu\sigma}(q)\,=0\tag{3.8}$$

is indeed satisfied to this order. In Ref. [4] the vacuum polarization was computed using a somewhat different expansion for the metric field, and a coordinate invariant expression for the one-loop counterterms was given in terms of operators quadratic in the curvature.

Finally, diagram 2h represents the contribution to the vacuum polarization due to one *massless* scalar particle,

$$-\frac{1}{20}\,G^2\,\frac{m_1 m_2}{\pi r^3}.\tag{3.9}$$

Its contribution to the vacuum polarization satisfies separately the Slavnov-Taylor identity, as one would expect from the covariant conservation law for the energy-momentum tensor associated with matter. Diagrams 2a and 2c do not give rise to any classical relativistic or quantum correction, while diagram 2f vanishes identically in dimensional regularization. Diagrams 2b, 2d, 2e, 2g and 2h give only quantum mechanical corrections, involving closed graviton loops in all cases, except 2b.

The sum of all contributions from diagrams 1a to 2g is therefore

$$+ \frac{1}{2} \, G^2 \, \frac{m_1 m_2 (m_1 + m_2)}{r^2} + \frac{122}{15} \, G^2 \, \frac{m_1 m_2}{\pi r^3} \qquad (3.10)$$

The contribution of $n$ species of massless scalar particles to the vacuum polarization (arising from diagram 2h) changes the quantum correction to the potential to

$$+ \frac{1}{60}(488 - 3n) \, G^2 \, \frac{m_1 m_2}{\pi r^3}, \qquad (3.11)$$

which represents a relatively small modification to the result for pure gravity if $n$ is small. Massless particles of higher spin will contribute additional terms to the vacuum polarization.

When the appropriate powers of $c$ and $\hbar$ are put back in, one obtains the following final answer for the corrected potential in pure gravity, valid to order $G^2$

$$V(r) \; = \; -G \, \frac{m_1 m_2}{r} \, \Big[ 1 - \frac{G(m_1 + m_2)}{2c^2 r} - \frac{122 G \hbar}{15 \pi c^3 r^2} \Big] \qquad (3.12)$$

As we alluded to previously, two very different length scales enter in the correction to the static Newtonian potential, namely the Schwarzschild radii of the heavy sources, $2Gm_i/c^2$, and the Planck length $(G\hbar/c^3)^{1/2}$. As a consequence there are two independent dimensionless parameters that appear in the correction term, involving the ratio of these two scales with respect to the distance $r$. Presumably the above calculation is meaningful only if these two length scales are much smaller than the distance $r$.

Our calculations are similar in spirit to the work of Ref. [6]. There the starting point is also a calculation of the scattering amplitude in the limit of small momentum transfer. The potential is defined there as the non-relativistic limit of the one particle reducible graphs in the crossed channel, which represents therefore a subset of the graphs considered here. We should point out that the results we obtain here are in complete agreement with the expected classical relativistic correction, as derived for example from the expansion of the Schwarzschild metric [20]. The sign of the quantum correction is found to be the same as in Ref. [6], and the magnitude of the correction is comparable. The sign of the quantum correction we obtain indicate that gravitational interactions increase (slowly) with distance, which shows similarities with the evolution of the coupling constant in pure Yang-Mills theories, but differs in sign from the QED radiative corrections to the static Coulomb potential. This result is also in agreement with the intuitive expectation that gravity couples universally to all forms of energy, and cannot be easily screened by quantum fluctuations.

Recently the authors of Ref. [10] have computed the corrections to the static Newtonian potential following the method of Ref. [21], thus extending to the next order in $G$ the calculation of Ref. [22]. In their work the radiative corrections to the potential are obtained by considering correlations between the action contributions from two heavy particle world lines, separated by a fixed geodesic distance. The results they obtain appear to correctly reproduce the classical relativistic correction, but arise from only a subset of two diagrams among the four which lead to the classical correction in Ref. [18]. In this last reference the ladder and crossed ladder diagrams give, using the same metric expansion, additional contributions which appear to be necessary in order to obtain the correct classical relativistic correction. These diagrams involve recoil of the massive particles, and have been neglected in the calculation of Ref. [10]. In our calculation we find that ladder and crossed ladder diagrams (1a and 1b), when carefully treated, contribute to the quantum correction. This probably explains why our results and the results of Ref. [10] differ in both sign and magnitude for the quantum correction.

Let us conclude by mentioning that we have little to say about what might happen to higher order in the perturbative expansion. In particular it is unclear if higher order corrections in $G$ can still lead to finite corrections in the long distance limit, as was found above to lowest non-trivial order. Whether higher derivative terms or string theory is needed to control the ultraviolet divergences appearing at higher loops remains an open question [23]. As we pointed out before, another important omission in the present calculation is represented by the absence of a cosmological constant term. This term substantially modifies the propagation properties of gravitons already at tree level, and leads to new, momentum independent, vertices and Feynman rules for gravitons which were not considered here.

Finally there is the issue of the non-perturbative definition of the Euclidean path integral for quantum gravity, which suffers from the problem of the unbounded fluctuations in the conformal mode, and for which an integration over complex conformal factors has been suggested, followed by an integration over conformal equivalence classes of metrics. In the framework of perturbation theory we did not have to deal with these difficult problems.

### Acknowledgements

One of the authors (HWH) would like to thank S. Deser, T.T. Wu and I. Muzinich for discussions, S. Vokos for correspondence, and the Theory Division at CERN for hospitality and support during the initial stages of this work.

# References

[1] R. P. Feynman, *Acta Phys. Polon.* **24** (1963) 697.

[2] R. P. Feynman, *"Lectures on Gravitation"*, Caltech Notes, edited by F. Morengo and W. Wagner (1963).

[3] B. DeWitt, *Phys. Rev.* **160** (1967) 1113; *Phys. Rev.* **162** (1967) 1195 and 1239.

[4] G. 't Hooft and M. Veltman, *Ann. Inst. H. Poincaré*, **20** (1974) 69.

[5] S. Deser and P. van Nieuwenhuizen, *Phys. Rev.* **D10** (1974) 401 and 410;
S. Deser, H. S. Tsao and P. van Nieuwenhuizen, *Phys. Lett.* **50B** (1974) 491;
*Phys. Rev.* **D10** (1974) 3337.

[6] J. F. Donoghue, *Phys. Rev. Lett.* **72** (1994) 2996; preprint UMHEP-408 (gr-qc/9405057).

[7] B. DeWitt and R. Utiyama, *J. Math. Phys.* **3** (1962) 608;
K. S. Stelle, *Phys. Rev.* **D16** (1977) 953.

[8] S. Weinberg, in *'General Relativity - An Einstein Centenary Survey'*, edited by S.W. Hawking and W. Israel, (Cambridge University Press, 1979).

[9] M. Roček and R. M. Williams, *Phys. Lett.* **104B** (1981) 31, and *Z. Phys.* **C21** (1984) 371.

[10] I. J. Muzinich and S. Vokos, preprint UW/PT-94-12 (December 1994).

[11] T. T. Wu, CERN Th, unpublished.

[12] D. M. Capper, G. Leibrandt, and M. Ramon Medrano, *Phys. Rev.* **D8** (1973) 4320.

[13] L. D. Fadeev and V. N. Popov, *Phys. Lett.* **25B** (1967) 29.

[14] E. S. Fradkin and I. V. Tyutin, *Phys. Rev.* **D2** (1970) 2841.

[15] M. Veltman, in *'Methods in Field Theory'*, Les Houches Lecture notes, Session XXVIII (North Holland, 1975).

[16] D. G. Boulware and S. Deser, *Ann. Phys.* **89** (1975) 193.

[17] M. J. Duff, *Phys. Rev.* **D7** (1972) 2317; *Phys. Rev.* **D9** (1974) 1837.

[18] Y. Iwasaki, *Progr. Theor. Phys.* **46** (1971) 1587.

[19] F. Radkowski, *Ann. Phys.* **56** (1970) 314.

[20] S. Weinberg, *'Gravitation and Cosmology'*, Ch. 9 (Wiley, New York 1972).

[21] G. Modanese, *Phys. Lett.* **325B** (1994), 354; *Nucl. Phys.* **B434** (1995) 697; preprint MIT-CTP-2217 (June 1993).

[22] H. Hamber and R. Williams, *Nucl. Phys.* **B435** (1995) 361.

[23] M. Goroff and A. Sagnotti, *Phys. Lett.* **160B** (1985) 81; *Nucl. Phys.* **B266** (1986) 709.